\documentclass[12pt,dvips,a4paper]{article}
\usepackage{a4wide}
\usepackage[dvips]{graphicx}
\usepackage{graphicx}
\usepackage{epsfig}
\usepackage{amsmath,amssymb,amsfonts}
\usepackage{cite}
\usepackage[x11names]{xcolor}

\graphicspath{{plots/}}
\DeclareGraphicsExtensions{.eps.gz,.eps,.ps,.ps.gz}

\parskip=\itemsep               
\newcommand{\lambdabar}{{\hbox{$\lambda_e$\kern-1.9ex\raise+0.45ex\hbox{--}
\kern+0.2ex}}}

\newcommand{\ii}{{\rm{i}}}
\newcommand{\ee}{{\rm{e}}}

\newif\ifhepph

\ifhepph\date{\empty}\fi

\title{
{\normalsize
\rightline{IPPP/07/28; DCPT/07/56; DESY 07-081}}\
\vskip 1cm
\bf\boldmath
Extending the reach of axion-photon regeneration experiments towards larger masses
with phase shift plates
       \vspace{21mm}}
\author{
Joerg Jaeckel\\[4mm]
Centre for Particle Theory, Durham University,\\ Durham, DH1 3LE, United Kingdom\\[8mm]
Andreas Ringwald\\[4mm]
Deutsches Elektronen-Synchrotron DESY, Notkestrasse 85,\\ D-22607 Hamburg, Germany\\[8mm]
}

\begin{document}
\begin{titlepage}
  \maketitle
\begin{abstract}
We present a scheme to extend the sensitivity of axion-photon regeneration experiments
towards larger masses with the help of
properly chosen and placed phase shift plates.
\end{abstract}


\thispagestyle{empty}
\end{titlepage}
\newpage \setcounter{page}{2}

Many proposals to embedd the standard model of particle physics into a more
general, unified framework predict a number of new very light particles which
are very weakly coupled to ordinary matter. Typically, such light particles
arise if there is a global continuous symmetry that is spontaneously broken
in the vacuum -- a notable example being the axion~\cite{Weinberg:1978ma,Wilczek:1977pj},
a pseudoscalar particle
arising from the breaking of a U(1) Peccei-Quinn symmetry~\cite{Peccei:1977hh} introduced to explain the
absence of CP violation in strong interactions. Other examples of light spin-zero bosons
beyond the standard model are
familons~\cite{Wilczek:1982rv}, Majorons~\cite{Chikashige:1980ui,Gelmini:1980re}, the dilaton, and moduli,
to name just a few.
We will call them axion-like particles, ALPs, in the following.

At low energies, the coupling of such an ALP, whose corresponding quantum field we denote by $\phi$,
to photons is described by an effective Lagrangian,
\begin{equation}
\label{em_pseudoscalar}
{\mathcal L} = -\frac{1}{4} F_{\mu\nu}F^{\mu\nu} + \frac{1}{2} \partial_\mu\phi\partial^\mu\phi
-\frac{1}{2}m_\phi^2\phi^2 -\frac{1}{4} g\phi F_{\mu\nu}{\tilde F}^{\mu\nu}
\,,
\end{equation}
where $F_{\mu\nu}$  (${\tilde F}_{\mu\nu}$) is the (dual) electromagnetic field strength
tensor\footnote{The effective Lagrangian~(\ref{em_pseudoscalar})
applies for a pseudoscalar ALP, i.e. a spin-zero boson with negative parity. In the case
of a scalar ALP, the $F_{\mu\nu}{\tilde F}^{\mu\nu}$ in Eq.~(\ref{em_pseudoscalar}) is replaced
by  $F_{\mu\nu}{F}^{\mu\nu}$. For the more general case where $\phi$ ceases to be
an eigenstate of parity\cite{Hill:1988bu}, see Ref.~\cite{Liao:2007nu}.}
and $m_\phi$ is the mass of the ALP. Correspondingly, in the presence of
an external magnetic field, a photon of energy $\omega$ may oscillate
into an ALP of small mass $m_\phi < \omega$, and vice versa~\cite{Sikivie:1983ip,Raffelt:1987im}.

\begin{figure}[h]
\begin{center}
\includegraphics*[width=10.cm,clip=]{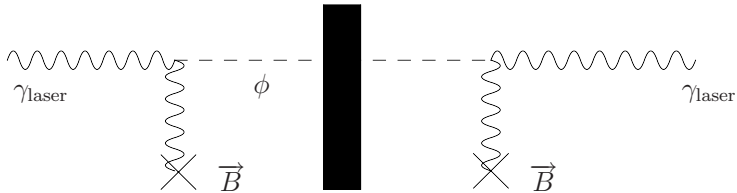}
\caption[...]{Schematic view of ALP production through photon conversion
in a magnetic field (left), subsequent travel through an optical barrier, and final
detection through photon regeneration (right).
\hfill
\label{fig:ph_reg}}
\end{center}
\end{figure}

The exploitation of this mechanism is the basic idea behind ALP-photon regeneration --
sometimes also called ``light shining through a wall'' --
experiments~\cite{Anselm:1986gz,Gasperini:1987da,VanBibber:1987rq} (cf. Fig.~\ref{fig:ph_reg}).
Namely, if a beam of photons is shone across a magnetic field, a fraction of these photons will turn
into ALPs. This ALP beam could then propagate freely through an optical barrier without being
absorbed, and finally another magnetic field located on the other side of the wall
could transform some of these ALPs into photons -- apparently regenerating these photons
out of nothing.

A pioneering experiment of this type was carried out in Brookhaven by the
Brookhaven-Fermilab-Rochester-Trieste (BFRT) collaboration, using two prototype
magnets for the Colliding Beam Accelerator~\cite{Ruoso:1992nx,Cameron:1993mr}.
Presently, there are worldwide several second generation ALP-photon regeneration experiments
under construction or serious consideration (cf. Table~\ref{tab:exp};
for a review, see Refs.~\cite{Ringwald:2006rf,Battesti:2007um}).
These efforts are partially motivated by the report from the PVLAS collaboration of
evidence for a non-zero apparent rotation of the polarization plane of a laser beam
after passage through a magnetic field~\cite{Zavattini:2005tm}. While the size of the observed effect greatly
exceeds the expectations from quantum electrodynamics~\cite{Adler:1971wn,Adler:2006zs,Biswas:2006cr},
it is compatible with the
expectations~\cite{Maiani:1986md}
arising in the context of a photon-ALP oscillation hypothesis.
Indeed, the rotation observed by PVLAS can be reconciled with the non-observation
of a signal by BFRT, if there exists an ALP with a mass $m_\phi\sim$~meV and a coupling
$g\sim 10^{-6}$~GeV$^{-1}$~\cite{Ahlers:2006iz}. Although these parameter values
seem to be in serious conflict with bounds coming from astrophysical considerations,
there are various ways to evade
them~\cite{Masso:2005ym,Jain:2005nh,Masso:2006gc,Jaeckel:2006xm,Mohapatra:2006pv,Jain:2006ki,Brax:2007ak}.
Therefore,
it is extremely important to check the ALP interpretation of PVLAS by purely laboratory
experiments~\cite{Jaeckel:2006xm}.
Moreover, it would be nice if in this way one might ultimately extend the laboratory search
for ALPS  to previously unexplored parameter values (see also Ref.~\cite{Sikivie:2007qm}).
In this letter, we propose
a method to extend the sensitivity of the planned photon-regeneration experiments
to higher ALP masses.

\begin{table}
\caption[]{Experimental parameters of upcoming photon regeneration experiments:
magnetic fields $B_i$ and their length $\ell_i$ on production ($i=1$) and
regeneration ($i=2$) side (cf. Fig.~\ref{fig:ph_reg}).}
\vspace{1.5ex}
\begin{center}
\begin{tabular}{|l|l|c||l||}
\hline
Name & Laboratory   & Magnets & Laser   \\
\hline
{\bf ALPS}\cite{Ehret:2007cm}&DESY/D     & $B_1=B_2=5$~T &
\\
 &            & $\ell_1=\ell_2=4.21$~m &  $\omega = 2.34$~eV  \\
\hline
{\bf BMV}\cite{Rizzo:Patras} &LULI/F        & $B_1=B_2=11$~T &
\\
 &        & $\ell_1=\ell_2=0.25$~m &  $\omega = 1.17$~eV  \\
\hline
{\bf LIPSS}\cite{Baker:Patras}&Jlab/USA    & $B_1=B_2=1.7$~T &
\\
 &    & $\ell_1=\ell_2=1$~m & $\omega = 1.17$~eV  \\
\hline
{\bf OSQAR}\cite{OSQAR}&CERN/CH    & $B_1=B_2=11$~T &
\\
 &        & $\ell_1=\ell_2=7$~m &  $\omega =1.17$~eV \\
\hline
 &    & $B_1=5$~T &
\\
{\bf PVLAS}\cite{Cantatore:Patras} & Legnaro/I     & $\ell_1=1$~m &   $\omega =1.17$~eV
\\
     &              & $B_2=2.2$~T & \\
&            &$\ell_2=0.5$~m  &\\
\hline
\end{tabular}
\end{center}
\label{tab:exp}
\end{table}

Let us start with an outline of the calculation of the photon $\to$ ALP conversion probability
$P_{\gamma\to\phi}$,
to lowest order in the coupling $g$. As emphasized in Ref.~\cite{VanBibber:1987rq}, this calculation
amounts to solving the classical field equations following from Eq.~(\ref{em_pseudoscalar}),
\begin{equation}
\label{em_pseudoscalar_fieldeq}
\partial_\mu F^{\mu\nu} = g \partial_\mu \left( \phi {\tilde F}^{\mu\nu}\right);
\hspace{6ex}
\left( \partial_\mu\partial^\mu + m_\phi^2\right) \phi = g {\vec E}\cdot {\vec B}
\,,
\end{equation}
to lowest order in $gB\ell$, where $\ell$ is the
linear dimension associated with the extent of the magnetic field\footnote{In the case of a
scalar ALP, the term ${\vec E}\cdot {\vec B}$ in Eqs.~(\ref{em_pseudoscalar_fieldeq}),
(\ref{lead_sol}), and (\ref{onedimensional}) is replaced by $\frac{1}{2}({\vec E}^2 - {\vec B}^2)$.}.
This can be done by
neglecting the modification of the electromagnetic field due to the presence
of the pseudoscalar field (through the right hand side of the first equation above).
Solving for $\phi$ in the second equation yields~\cite{Sikivie:1983ip,VanBibber:1987rq}
\begin{equation}
\label{lead_sol}
\phi^{(\pm)}(\vec x,t) = {\rm e}^{-{\rm i}\omega t}
\int {\rm d}^3x^\prime\frac{1}{4\pi}
\frac{{\rm e}^{\pm {\rm i}k_\phi |\vec x - {\vec x}^\prime|}}{|\vec x - {\vec x}^\prime|}
g \vec E({\vec x}^\prime )\cdot \vec B({\vec x}^\prime )
\,,
\end{equation}
where the energy $\omega$ and the modulus of the three-momentum
$k_\phi$ are related by $k_\phi = \sqrt{\omega^2 -m_\phi^2}$.
\begin{figure}
\begin{center}
\includegraphics*[bbllx=26,bblly=222,bburx=583,bbury=610,width=.5\textwidth]{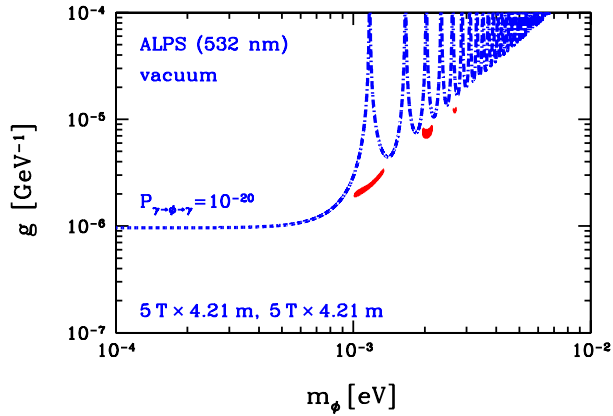}
\caption[...]{Two photon coupling $g$ of the (pseudo-)scalar versus its mass $m_\phi$.
Iso-contour of the regeneration probability
$P_{\gamma\to\phi\to\gamma}=P_{\gamma\to\phi}P_{\phi\to\gamma}$,
for the parameters of the ALPS experiment, i.e. magnetic fields $B_1=B_2=5$~T, over a length
$\ell_1=\ell_2=4.21$~m, exploiting a green ($\lambda =532$~nm)
photon beam, corresponding
to $\omega = 2.34$~eV, in vacuum.
Also shown in red are the 5 sigma allowed regions~\cite{Ahlers:2006iz}
from PVLAS data on rotation~\cite{Zavattini:2005tm} plus BFRT data on rotation,
ellipticity, and regeneration~\cite{Cameron:1993mr} plus Q\&A data on rotation~\cite{Chen:2006cd}.
\hfill
\label{regprop_stan}}
\end{center}
\end{figure}
This solution simplifies even more if we specialize to the usual experimental configuration
of a laser photon beam send along the $x$-axis with fixed linear polarization
in the $z$ direction.
If the transverse extent of the magnetic field is much larger than that of the laser beam, the problem
is effectively one-dimensional.
In one dimension and taking into account only ALPs that propagate into the positive $x$-direction,
Eq.~\eqref{lead_sol} becomes,
\begin{equation}
\label{onedimensional}
\phi^{(+)}(x,t)=\ee^{-\ii(\omega t -k_{\phi}x)}\frac{\ii g}{2k_{\phi}}\int {\rm d}x^{\prime}\,
\vec E(x^{\prime})\cdot \vec B(x^{\prime}).
\end{equation}
Inserting in Eq.~(\ref{onedimensional}) furthermore the appropriate plane wave form
${\vec E}_0(\vec x,t )={\vec e}_z E_0 {\rm e}^{{\rm i}\omega (x-t)}$
for the electric field of the laser beam and assuming, as realized in all the proposed
experiments, a magnetic field
with fixed direction along the $z$-axis and possibly variable (as a function of $x$)
magnitude, ${\vec B}_0({\vec x}) ={\vec e}_z B_0(x)$, one ends up with the solution\footnote{The
solution~(\ref{sol_stand}) applies also in the case of a scalar ALP, if the magnetic field
direction is chosen to point into the $y$ direction, ${\vec B}_0({\vec x}) ={\vec e}_y B_0(x)$ (or,
alternatively, if the polarization of the laser is chosen to point in the $y$ direction).}
\begin{equation}
\label{sol_stand}
\phi^{(\pm)}(\vec x,t) = \frac{{\rm i}g}{2k_\phi} E_0\, {\rm e}^{-{\rm i}(\omega t -k_\phi x)}
\int {\rm d}x^{\prime}\, {\rm e}^{{\rm i}qx^{\prime}} B_0(x^{\prime})
\,,
\end{equation}
where
\begin{equation}
\label{mom_trans}
q=k_\gamma - k_\phi
=\omega - \sqrt{\omega^2 - m_\phi^2} \approx \frac{m_\phi^2}{2\omega}
\end{equation}
is the momentum transfer to the magnetic field, i.e. the modulus of the
momentum difference between the photon and
the ALP.
The probability that a photon converts into
an axion-like particle and vice versa
can be read off from Eq.~(\ref{sol_stand}) and reads~\cite{Sikivie:1983ip,VanBibber:1987rq}
\begin{equation}
\label{prop_stan}
P_{\gamma\to\phi} = P_{\phi\to\gamma } = \frac{1}{4}\, \frac{\omega}{k_\phi}\,
g^2\,\left|\int {\rm d}x^\prime\,{\rm e}^{{\rm i}qx^\prime}\,
B_0(x^\prime)\right|^2 \,,
\end{equation}
which reduces, for a constant magnetic field, $B_0(x^\prime )={\rm const}$, of linear extension
$\ell$, to
\begin{equation}
\label{sinsuppress}
P_{\gamma\to\phi} \approx
g^2\,B_0^2\,\sin^2\left( q\ell/2 \right)/q^2\,.
\end{equation}

Clearly, in the experimental setup considered, the maximum conversion probability,
$P_{\gamma\to\phi} \approx g^2 B_0^2 \ell^2$, is attained at small momentum transfer,
$q=m_\phi^2/(2\omega)\ll 1$, corresponding to a small ALP mass.
For this mass range, the best limits are obtained in a straightforward manner by exploiting strong
and long dipole magnets, as they are used for storage rings such as HERA~\cite{Ringwald:2003ns}
or LHC~\cite{Pugnat:2005nk}, cf. the experiments ALPS~\cite{Ehret:2007cm} and
OSQAR~\cite{OSQAR}, respectively (see Table~\ref{tab:exp}). However, for larger masses,
the sensitivity of this setup rapidly diminishes.

We illustrate this in Fig.~\ref{regprop_stan}, which displays an iso-contour of the light shining
through a wall probability in the $g$-$m_\phi$ plane, exploiting the experimental parameters of the
ALPS experiment~\cite{Ehret:2007cm}.
Clearly, for this setup, the parameter region in $g$ vs. $m_\phi$ suggested
by the combination of BFRT plus Q\&A exclusion and PVLAS evidence can not be probed. This is even
more dramatic for the OSQAR experiment, which exploits an LHC magnet.
Moreover, increasing the refraction index by filling in
buffer gas does not help since it works in the wrong direction (contrary to the
claim\footnote{In a refractive medium, the laser beam has a phase velocity
$1/n\equiv v \equiv \omega/k_\gamma$. The momentum transfer~(\ref{mom_trans}) reads then
$q=n\,\omega - \sqrt{\omega^2 - m_\phi^2} \approx
\frac{m_\phi^2}{2\omega}+(n-1)\,\omega$.
The second term in this expression has the opposite sign
as the corresponding term in Ref.~\cite{Ehret:2007cm}. Correspondingly, one would need a buffer gas
with refraction index less than unity, i.e. a plasma, in order to decrease $q$ (and thereby
maximize the conversion probability~(\ref{sinsuppress})) rather than to increase it.
A.R. would like to thank Aaron Chou for pointing out
the correct sign. } in the
ALPS letter of intent~\cite{Ehret:2007cm}).

\begin{figure}
\begin{center}
\includegraphics*[bbllx=26,bblly=222,bburx=583,bbury=610,width=.5\textwidth]{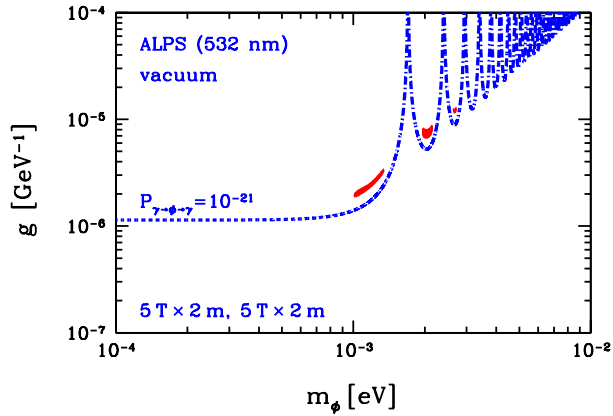}
\caption[...]{Iso-contour of the regeneration probability, as in Fig.~\ref{regprop_stan}, but with reduced
lengths of the magnetic field region.
Note, that the regeneration probability is reduced by a factor of $10$.
\hfill
\label{regprop_short}}
\end{center}
\end{figure}

A simple possibility to probe the meV region\footnote{Another
possibility to probe larger ALP masses even with a long magnet would
be to exploit VUV or X-ray free-electron laser
beams~\cite{Ringwald:2001cp,Rabadan:2005dm,Kotz:2006bw}. However, at the
moment conventional lasers seem to offer better prospects (see also
Ref.~\cite{Sikivie:2007qm})} in the ALPS setup is to reduce the
effective length of the magnetic field region both on the production
and detection side of the magnet by shortening the beam pipe on both
sides. As can be seen in Fig.~\ref{regprop_short}, this possibility
enables to extend the mass region probed by the experiment, however
at the expense of sensitivity: one looses about one order of
magnitude in the light shining through a wall probability.

Another idea to extend the sensitivity towards larger ALP masses was introduced in
Ref.~\cite{VanBibber:1987rq}. There,  it was shown that a segmentation
of the magnetic field into regions of alternating polarity  gives a form factor
$\int {\rm d}x^\prime \exp ({\rm i}qx^\prime) B_0(x^\prime)$
that peaks at a nonzero value of $q$, thereby giving
sensitivity to higher-mass pseudoscalars. In fact, the conversion probability~(\ref{prop_stan})
reads~\cite{VanBibber:1987rq,Afanasev:2006cv},
in a magnet with $N$ segments of alternating field direction (but the same magnitude $B_0$),
\begin{eqnarray}
P_{\gamma\to\phi} &\approx &
g^2\,B_0^2\,\frac{\sin^2\left( qd/2 \right)}{q^2}
\left|\sum_{k=1}^{N}(-1)^k \exp\{{\rm i}(2k-1) qd/2\}\right|^2
\\[1.5ex] \nonumber &=&
\frac{g^2\,B_0^2}{q^2}
\left\{
\begin{array}{clc}
\sin^2\left( q\ell/2 \right) \tan^2\left( q\ell/(2N) \right)& {\rm for} &
N\ {\rm even} \\
\cos^2\left( q\ell/2 \right) \tan^2\left( q\ell/(2N) \right)& {\rm for} & N\ {\rm odd}
\end{array}
\right.
\,,
\end{eqnarray}
where $d=\ell/N$ is the length of each of the
$N$ segments. For $N>1$, this indeed gives rise to more sensitivity at non-zero values of $q$.

\begin{figure}
\begin{center}
\includegraphics*[bbllx=26,bblly=222,bburx=583,bbury=610,width=.5\textwidth]{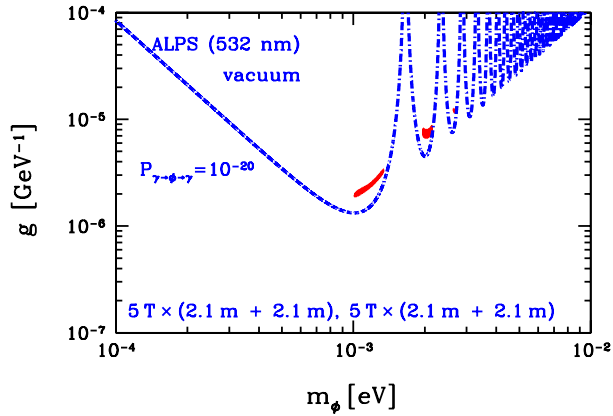}
\caption[...]{Iso-contour of the regeneration probability, as in Fig.~\ref{regprop_stan}.
Here, we used one phase shift (``$\lambda/2$'')
plate each in the middle of the generation and the regeneration sides.
\hfill
\label{regprop_ret}}
\end{center}
\end{figure}

In this letter, we will introduce a similar, but more practical possibility based on the use of
phase shift plates.
The idea is very simple. From our starting point,
Eq.~(\ref{onedimensional}), we can see that what counts is actually $\vec E(x^\prime )\cdot \vec B(x^\prime)$.
The configuration based on $N$ alternating magnetic fields is therefore equivalent
to a configuration with non-alternating magnetic field, however with $N-1$ retardation
plates with phase shift $\pi$ (``$\lambda/2$'' plates) inverting the sign of the electric field,
placed equidistantly over the length $\ell$ of the magnet.
In this case we have alternating signs of $\cos\theta$, where $\theta$ is the angle between $\vec E$
and $\vec B$,
instead of alternating signs of the magnetic field.  But both cases have an identical profile of
$\vec E(x^\prime )\cdot \vec B(x^\prime)$.
In Fig.~\ref{regprop_ret}, we show that with a proper choice of the number and
positions of such phase shifters, ALPS should easily cover the
region of parameter space suggested by PVLAS + BFRT + Q\&A. The same applies for OSQAR.

Let us now get a more intuitive understanding of how this works and see how we can do even better.
The crucial part in Eq. \eqref{sol_stand} is the integral
\begin{equation}
\label{form}
f(q)=\int {\rm d}x^{\prime} \ee^{{\rm i}qx^{\prime}}B_0(x^{\prime}).
\end{equation}

For a constant magnetic field of length $\ell$ the oscillating factor $\ee^{\ii qx^\prime}$
suppresses the integral
compared to the massless case
with $q=0$, where the integral is simply
\begin{equation}
\label{formmax}
|f(q)|<|f(0)|=\left|\int^{\ell}_{0} {\rm d}x^{\prime} B_{0}\right| =
B_0\ell, \quad{\rm{for}}\,\,B_0(x)={\rm{constant}}.
\end{equation}
This suppression arises because coherent production of ALPs works only if the ALP and the photon are in phase.
The factor $\ee^{\ii qx^\prime }$ accounts for the phase difference between ALP and photon.

\begin{figure}
\begin{center}
\includegraphics*[width=.65\textwidth]{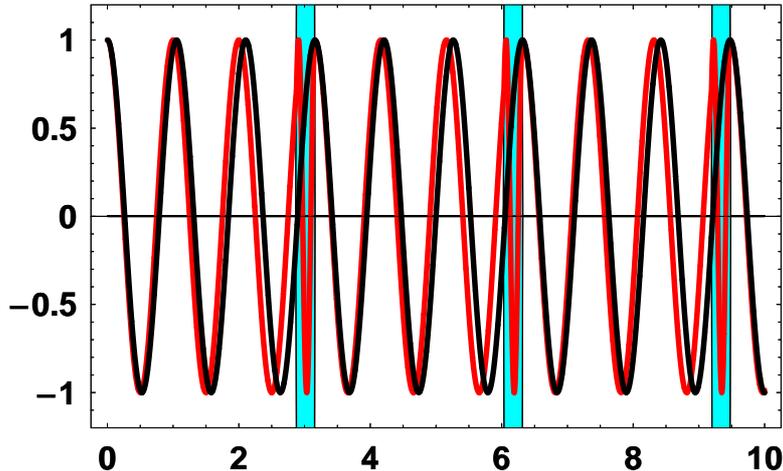}
\caption[...]{Illustration of the effect of a properly chosen and placed
phase shift plate on the phase relation between photon and ALP (this simplified picture shows
only the phase relation; the amplitudes of photon and ALP
are not correct in this picture). Photon (red) and ALP (black) start in phase. Due to their different wavelength
they are, however, somewhat out of phase after
several oscillations - say by an amount $\zeta$.
This is corrected by introduction of a phase shift plate that causes the photon to get an extra
phase $2\pi-\zeta$. In other words the plate causes the photon to complete the extra wiggle.
\hfill
\label{illu}}
\end{center}
\end{figure}

To improve the situation one would want to bring photon and ALP back into phase with each other. This can be
achieved in a simple way by the introduction of
phase shift plates. A simplified picture of the phase correction process is given in Fig. \ref{illu}.
At the beginning,
photon (red) and ALP (black) are in phase. However, due to its mass, the ALP has a slightly larger wavelength
than the photon. After a few oscillations photon and
ALP are more and more out of phase. Then we insert the phase shift plate (turquoise). With refractive
indices $n>1$ we cannot make the wavelength larger.
So it is not possible to ``delay'' the photon until the ALP has caught up. What we can do, however, is to is
to \emph{increase} the phase difference between ALP
and photon such that it is exactly $2\pi$ (the photon does an extra wiggle in Fig. \ref{illu}.).
Now, a phase shift of $2\pi$ is exactly equivalent to a phase shift of $0$. Photon and ALP are in phase again.
Therefore, we can keep photon and ALP in phase
over quite long distances simply by inserting a suitable phase shift plate whenever the phase difference
 becomes too large, and we get coherent production over
the whole length of the magnet.

Let us now understand more quantitatively how this works.
To derive Eqs. \eqref{sol_stand} and \eqref{form} we have assumed that the photon is a plane wave. Therefore,
we can identify $qx^\prime =(k_{\gamma}-k_{\phi})x^{\prime}$
as the phase difference between the photon wave and the ALP wave at the point $x^{\prime}$. In general we
should write (cf. Eq. \eqref{onedimensional})
\begin{equation}
\label{phases}
f(q)=\int {\rm d}x^{\prime} \ee^{i (\varphi_{\gamma}(x^{\prime})-\varphi_{\phi}(x^{\prime}))}B_0(x^{\prime})
\,,
\end{equation}
where $\varphi_{\gamma,\phi}$ are the phases of the photon and the ALP fields, respectively.

Let us imagine a situation where we insert $N-1$ thin\footnote{One might ask what happens to the photon-ALP
system
inside the material of the plate. One can check (cf., e.g., Ref.~\cite{Raffelt:1987im})
that for sufficiently large refractive index of the material, $n-1\gg m^2_{\phi}/(2\omega^2)$, the mixing
between photon and ALP is effectively switched off
compared to the mixing in vacuum. Photon and ALP simply propagate through the plate without changing their
amplitudes (the phases change, of course).
In other words, the thickness of the plates has to be subtracted from the total length of the production or
regeneration region.
That is why we require thin plates. For practical purposes, this is a rather mild constraint.
For $n-1\sim 0.1$,
the thickness of the plates required for a phase shift of the order of $2\pi$ is only
$d\sim 10\, \lambda\sim 10\,\mu m$, which is tiny
compared to the typical lengths of the production/regeneration regions which
are of the order of a few~m.}, non-reflective\footnote{Reflected photons are effectively lost.} plates that
accelerate the photon phase by $\kappa$ at
equidistant
places $s\ell/N=s\Delta x$, $s=1\ldots N-1$, in a constant magnetic field of length $\ell$.
The plates affect only the photon. The ALP phase remains unaffected. Therefore, we have,
\begin{eqnarray}
\varphi_{\gamma}(x)\!\!&=&\!\! k_{\gamma}x+s\kappa \quad{\rm{for}}\quad s\Delta x<x\leq (s+1)\Delta x,
\quad\quad\Delta x=\frac{\ell}{N},
\\\nonumber
\varphi_{\phi}(x)\!\!&=&\!\!k_{\phi} x.
\end{eqnarray}
Inserting this into Eq. \eqref{phases}, we find
\begin{eqnarray}
\label{phasecorrected}
f(q)\!\!&=&\!\!B_{0} \sum^{N-1}_{s-0}\int^{(s+1)\Delta x}_{s\Delta x}
{\rm d}x^{\prime} \ee^{\ii(qx^{\prime}+s\kappa)}
=B_{0}\ee^{\frac{\ii}{2}(q\ell+(N-1)\alpha)}\frac{2\ii \sin\left(\frac{q\Delta x}{2}\right)}{q}
\frac{\sin\left(\frac{N}{2}(q\Delta x+\kappa)\right)}{\sin\left(\frac{1}{2}(q\Delta x+\kappa)\right)}.
\end{eqnarray}
We can now choose the number of plates $N$ and the phase shift $\kappa$ according to the recipe described above.
First we choose
$N$ large enough such that
\begin{equation}
\frac{1}{2}q \Delta x\ll1
\end{equation}
And then we choose $\kappa$ such that the phase difference that has accumulated over $\Delta x$ is
``completed'' to $2\pi$,
\begin{equation}
\label{optimal}
\kappa=2\pi-\frac{1}{2}q\Delta x.
\end{equation}
Evaluating Eq. \eqref{phasecorrected} in the limit $N(q \Delta x+\kappa)/2\to 0$ one finds
\begin{equation}
|f(q)|=B_{0}\Delta x\, N=B_{0}\ell.
\end{equation}
And we have coherent production over the whole length $\ell$.

\begin{figure}[t]
\begin{center}
\includegraphics*[bbllx=26,bblly=222,bburx=583,bbury=610,width=.49\textwidth]{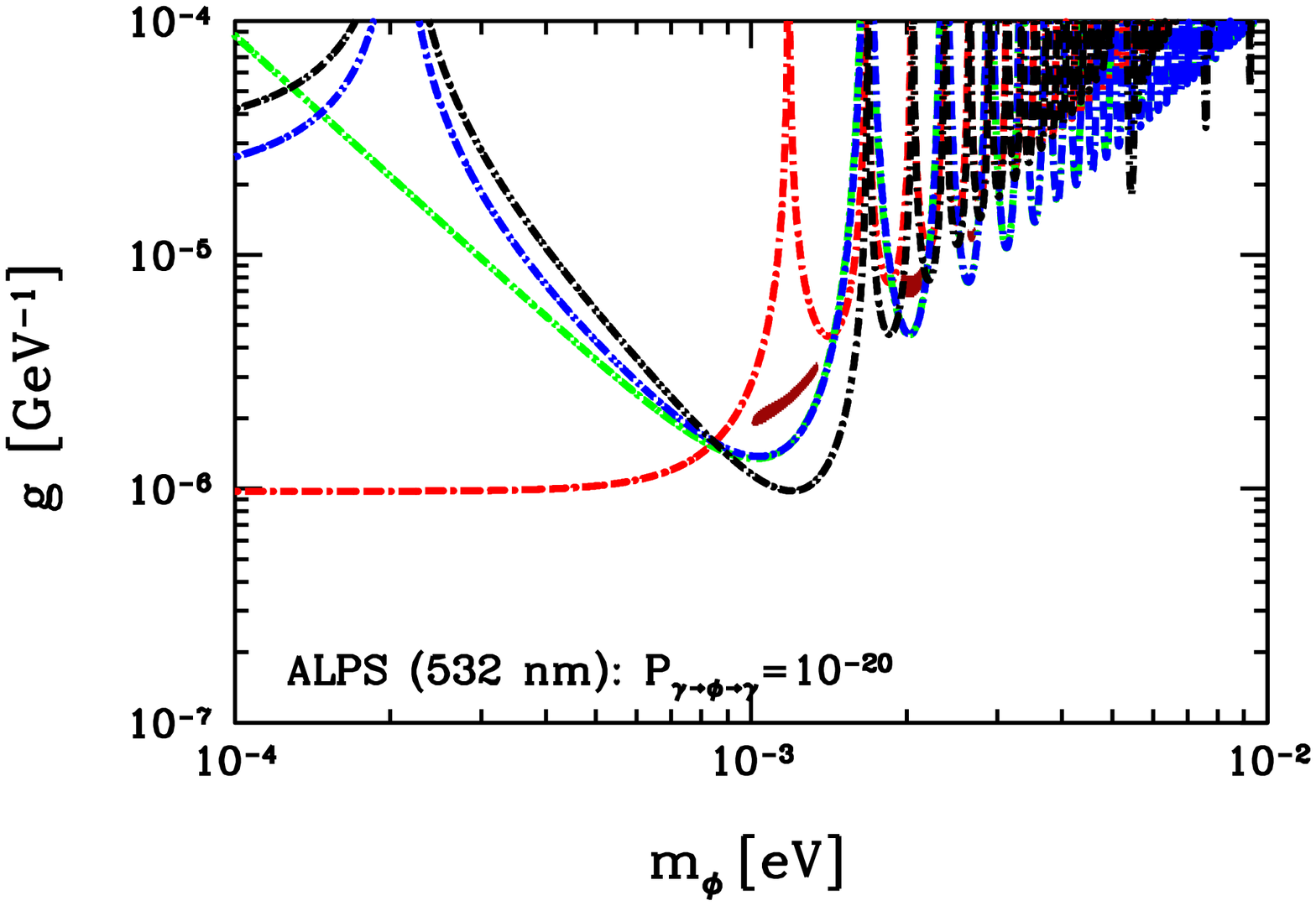}
\hfill
\includegraphics*[bbllx=26,bblly=222,bburx=583,bbury=610,width=.49\textwidth]{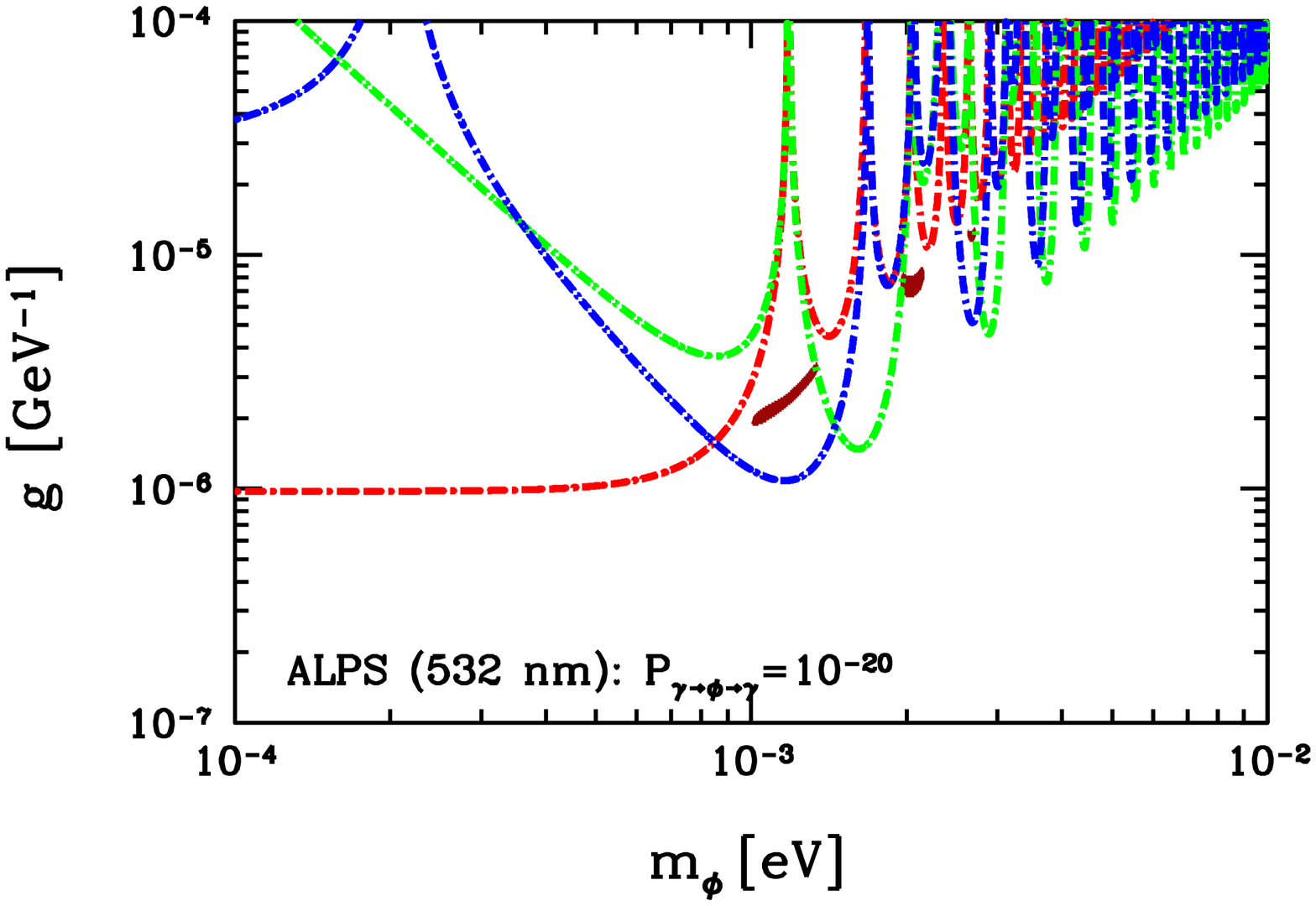}
\caption[...]{
Iso-contours of the regeneration probability, as in Fig.~\ref{regprop_stan}.
In the left figure, we have used no phase correction (red), one plate with $\kappa=\pi$ (green), and
one plate with the optimal choice of $\kappa$ according to Eq. \eqref{optimal} for $m_{\phi}=1.2\,{\rm{meV}}$
(blue).
The black curve is for 20 plates with the optimal choice
of $\kappa$.
In the right figure, we have the same but with $3$ plates for the green and blue curves.
\hfill
\label{optim1}}
\end{center}
\end{figure}

The potential of this approach is demonstrated in Fig. \ref{optim1}
for the example of the ALPS experiment. In the optimized mass region
we get more than ten times as many regenerated photons as we would
get if the length of the magnet is reduced as in
Fig.~\ref{regprop_short}.

\begin{figure}[t]
\begin{center}
\includegraphics*[bbllx=26,bblly=222,bburx=583,bbury=610,width=.5\textwidth]{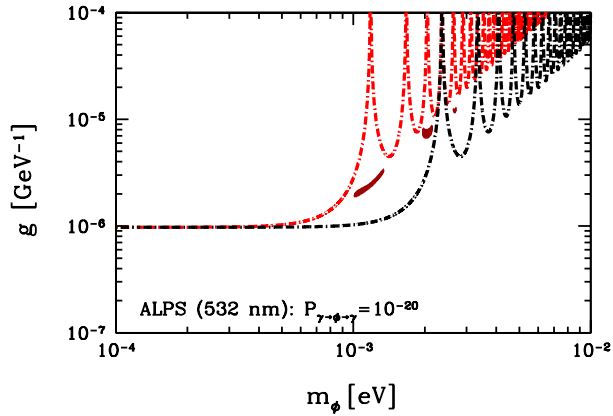}
\caption[...]{
Iso-contours of the regeneration probability, as in Fig.~\ref{regprop_stan}.
This figure demonstrates the potential for scanning through a whole range of masses by choosing the right
$\kappa$ for each $m_{\phi}$. The red curve
is the sensitivity without phase correction. The black curve is obtained by using three plates but scanning
through a whole range of $\kappa$. In other
words to obtain this curve one would insert the plates. Measure. Change the plates to a slightly different
value of $\kappa$ and measure again. This is repeated
for all values of $\kappa$ in the range~$[0,2\pi]$.
\hfill
\label{optim2}}
\end{center}
\end{figure}

Another practical advantage of this method is that we can scan through a whole mass range. Performing several
measurements with different phase shift plates
we can always choose for each
$q$, i.e. for each $m_{\phi}$, plates with an appropriate
$\kappa$ such that it is close enough to its optimal value~\eqref{optimal},
\begin{equation}
\label{optimalchoice}
\frac{1}{2}|q(m_{\phi})\ell-N\kappa|\ll 1.
\end{equation}
For an infinite number of plates this would allow to extend the mass range all the way to the frequency
$\omega$ of the photons\footnote{Above the photon frequency,
ALP production is energetically forbidden and $q$ becomes imaginary.}.
In practice, we can insert only a finite number of phase shift plates and Eq.~\eqref{optimalchoice} cannot be
fulfilled for too large masses.
But, already a small number of plates leads to a remarkable increase of the sensitivity for higher masses, as
we can see from Fig. \ref{optim2}.

{\emph{In summary:}} So called ``light shining through a wall'' experiments are a promising tool to search for
light particles
coupled to photons.
In this note we have shown how the reach of
such an experiment can be extended towards larger masses by inserting properly chosen phase shift plates.
Although our explicit discussion is for the case of spin-0 axion-like particles the method works in general
for particles exhibiting photon-particle-photon
oscillations.

\section*{Acknowledgments}
We would like to thank Giovanni Cantatore, Aaron Chou, Marin Karuza, Axel Lindner, Giuseppe Ruoso,
Pierre Sikivie, and Karl van Bibber for interesting discussions.

\end{document}